# What Attracts Employees to Work Onsite in Times of Increased Remote Working?

Darja Smite[1,2], Eriks Klotins[1], Nils Brede Moe[2,1]

[1] Blekinge Institute of Technology, Sweden; [2] SINTEF, Norway

***Abstract***: COVID-19 pandemic has irreversibly changed the attitude towards office presence. While previously remote workers were met with skepticism and distrust, today the same applies to companies prohibiting remote working. Albeit many workspaces are half empty. In this paper, we offer insights into the role of the office, corporate policies and actions regarding remote work in eight companies: Ericsson, Knowit, SpareBank 1 Utvikling, Spotify, Storebrand, Telenor, Company-X, Company-Y, and their sites in Sweden, Norway and the UK. Our findings are twofold. First, we found that companies indeed struggle with office presence and a large share of corporate space (35-67%) is underutilized. Second, we found that the main motivator for office presence is Connection and community, followed by Material offerings, Preference and Duty. Finally, we summarize actionable advice to promote onsite work, which is likely to help many other companies to rejuvenate life in their offices.

## Introduction

The Covid-19 pandemic demonstrated that working from home (WFH) is a feasible alternative to office work. Due to better-than-expected WFH experiences and the change in attitude towards remote workers [1], many engineers prefer WFH flexibility over commuting to the office [2, 3]. The call for increased flexibility at the workplace resulted in many companies experimenting and adjusting their corporate policies. Companies like Atlassian, Slack, Square, and Salesforce, announced that engineers have the freedom to decide where to work and encouraged to work remotely. New corporate identities lie on a continuum between office-first, hybrid, remote-first or fully remote arrangements [2].

Emerging hybrid work habits raised a considerable debate and divided employees, employers, and the public into two fronts: proponents of remote working and proponents of office work. This debate often opposes two key values: the freedom of individual choices for where to perform the work best (for the former group) and collaboration and serendipitous encounters (for the latter group).

On one hand, WFH is associated with higher job satisfaction and commitment [12, 13], and helps to save costs. Companies as Yelp increase the amount of remote work and become all-remote, abandone office buildings and keep one headquarters office in a "hoteling" mode [4]. Empty offices indeed make many employers reconsider their willingness to pay expensive rent and motivate downsizing decisions.

On another hand, increased remote working is also associated with social and professional isolation, and hindered networking, informal learning and mentoring [12]. Some even suggest that the benefits of autonomy when WFH can be outweighed by the collapsing boundaries between work and non-work spheres and resulting overwork (known as the flexibility paradox) [12]. In-person interactions are also found irreplaceable for ideation, brainstorming, and thus innovation [7]. Companies driven by these fears thus insist on mandatory office presence, such as Apple and Google with three days per week [5], or Twitter's full-time office work [6]. As we describe in the article, most of our case companies too fear the negative effects of WFH and associate office presence with efficient collaboration and mutual help, sense of community, innovation and creativity, and the main promoter of corporate culture.

Finally, many employers do not necessarily embrace remote working but are not ready to force employees back, fearing resignations [2]. This is especially true in Scandinavia, where the willingness to preserve democratic leadership are put in conflict with the willingness to preserve a highly collaborative and innovative work culture.

We can conclude that there is a growing consensus that remote working is an important alternative work arrangement, although with dual consequences [13]. In other words, we no longer have a choice – the future of work is increasingly hybrid. Therefore, we need to better understand emerging needs of the employees and the changing role of the office, which are in the focus of this article.

## Our study

With our study, we aim to better understand the mechanisms that gravitate employees towards the office, such as compensation for the deficiencies of the home office or the unavailability of certain functions, and the requirements for the modern workplace to inform decisions around renovations and work policies.

We conducted a multi-case study of motivations to work in the office in eight companies developing IT solutions. Case selection was based on convenience sampling. Notably, in addition to cases from Sweden and Norway, we included the UK site of a Swedish company, which extends the Nordic context. Analysis shows that findings from the UK case are consonant with those from the Nordic locations.

For data collection, we employed a non-intrusive, opt-in approach. We solicited the reasons for office presence using a poster with empty post-it notes and a catchy call:

> "You would not believe why your colleagues came to the office today. What brought YOU to the office?"

To maximize participation, posters were put up in communal areas where software teams reside. With this approach we avoid biasing responses towards neutral or socially acceptable answers if asked directly, or towards predefined response options. Posters were available during several workdays and collected 271 post-its, some of which listed several reasons, some were unclear. Unclear notes that we failed to clarify, were discarded, leaving 282 valid statements.

For analysis, we used a combination of structural, provisional, and in-vivo coding [8]. In the first pass, we read all notes, and identified if they contained one or multiple statements (structural coding) and what the recurring statements were across the notes (in-vivo coding). From the recurring statements we identified the main categories for provisional coding in several cycles. For example, a post-it note that mentioned "Eating waffles" and "Waffles" (in-vivo coding) were merged as "Waffles" (first cycle of provisional coding), and then merged into larger categories "Good lunch options, waffles" (second cycle) and "Food and drinks" (third cycle). All emerging categories are summarized in Figure 1.

To deepen our analysis, we collected quantitative office presence data (if available), visited the offices and conducted interviews. Office presence data was collected from the access card logs; we received summarized information for selected days and/or weeks. Office visits were used to observe the workplace and office occupancy on specific days. During the visits we kept a diary, used in describing the company offices. During the interviews, we captured company contexts, current situation and managers' perspectives on office presence, changes in the workplace, and office perks. Case narratives were validated by the managers.

Collected data is summarized in Table 1.

Our study is subject to several limitations. Collecting post-its during less-busy days might bias to result in specific reasons. For example, people who work in the office on Fridays (more empty days) might be driven by the unwillingness to work at home and not by the socialization or collaboration opportunities. We have addressed this by organizing the post-it collection during several days and, in some cases, different periods, as well as present aggregated results.

**Table 1:** Overview of the data

| Company | Site, Location | Presence data | Work policy | Response notes | Interviews, group sessions |
|---|---|---|---|---|---|
| Ericsson | Karlskrona, Sweden | Access cards (w39 2022-w7 2023) | X | 8 (Dec 19-20, 2022; Jun 23-30, 2023) | Interview with a unit manager; Written interview with a communications manager, Group discussion in the leadership forum |
| Knowit | Oslo, Norway | Qualitative analysis | X | 26 (Dec 16&19, 2022) | Written interview with the head of R&D |
| Company-X * | Malmö, Sweden | Desk occupancy (w47 2022-w7 2023) | N/A | 26 (Jun 13, 2023) | Written interview with the head of the startup acceleration program |
| Company-Y * | Lund, Sweden | Access cards (w47 2022-w7 2023) | N/A | 29 (Dec 19, 2022) | Interview with a strategic partnership manager; Group discussion with five team managers |
| SpareBank 1 Utvikling (SB1U) | Oslo, Norway | Access cards (w6-w7 2023) | X | 15 (Jan 18-23, 2023) | |
| Spotify | London, the UK | N/A | X | 23 (Dec 15-16, 2022) | Interview with a data engineering manager; Informal discussion with a product manager |
| | Stockholm, Sweden | N/A | X | 37 (Dec 19-20, 2022) | Interview with an engineering manager |
| Storebrand | Oslo, Norway | Access cards (w6-w7 2023) | X | 47 (Jan 19-20, 2023) | Interviews with a HR manager |
| Telenor | Karlskrona, Sweden | Aggregated weekday office presence | X | 17 (Dec 20, 2022) | Interview with an IT leader |
| | Oslo, Norway | N/A | X | 43 (Jan 11-13, Jun 26-30, 2023) | Group discussion in the research unit |

\* Pseudonyms, based on the companies' wish to remain anonymous

## Eight Stories

Table 2 summarizes corporate policies, office occupancy, renovation projects, office perks and activities that impact office presence.

### Ericsson

Ericsson is a major multinational networking and telecommunications company that employs over 100'000 employees worldwide and delivers complex software-intensive systems and services to telecommunications service providers, and enterprises. We studied one Swedish site of Ericsson with around 700 employees.

After the pandemic, the company introduced remote work rules, which demand employees to work from the office *"at least 50% of the time during a calendar year"*. However, office presence is in the spotlight of corporate and local site managers, who repeatedly emphasize the importance of helping and supporting colleagues over focusing entirely on one's own tasks, as well as innovating and driving the corporate culture.

Despite the management call to increase office presence, the office remains more than half-empty, even in a small town. Underutilized services have been thus looked upon as an opportunity to cut the costs. Thus far, there have not been any downsizing (or renovations) in the office, and there is still space for everyone. But there are plans to cut the space when the office lease will be renegotiated.

To increase office presence, Ericsson management and employees themselves organize numerous office-based activities. Teams organize joint breakfasts, management organizes week-long "return to the office" events, onsite seminars, gatherings and afterworks. The effectiveness of these activities has already surfaced in that around 13% employees work more on site than remotely.

### Knowit

Knowit is an IT consultancy company with a strong presence in the Nordic countries as well as other parts of Europe. This study focuses specifically on Knowit Objectnet, a subsidiary with approximately 175 consultants, and its main office in Oslo. In 2022, Knowit moved to a brand-new office in downtown Oslo, the capital of Norway.

The majority of consultants work directly with clients, often on-site at the client's location. However, they also have the option to work remotely from home or from the newly renovated main office in Oslo, which officially opened in the spring of 2022. The employees are allowed to make these decisions autonomously, taking customer needs into account. The main office has intentionally been designed with fewer workstations than the total number of employees, aiming to serve as a space for physical meetings, work-related activities, and fostering social connections among colleagues. The new office no longer provides free parking.

The employees work in teams and, when deciding whether to work from home or onsite, they are expected to prioritize team's efficiency over individual efficiency. Management clearly communicates their preference for employees to be present in the office, actively contributing to the work environment. This emphasis on office presence is driven by their commitment to building a strong company culture and fostering a sense of community, which is perceived as more challenging to achieve when everyone works remotely.

### Company-X

Company-X is an innovative branch of Company-Y employing around 40 engineers and a large network of Company-Y experts behind who develop and deliver a complete smart office solution for customers in various industries.

As an innovative company, Company-X values office presence and considers it as a prerequisite for creativity. However, there is no formally required presence, as the company management is against forcing onsite work.

During the pandemic, Company-X moved into a new office and focused on creating an attractive workplace that people would like to go to. The office is easy to reach and has a 360-degree panorama view from the 24[th] floor over the south Sweden and the coast of Denmark, as well as a balance of work areas and socialization areas. Besides, to foster social gatherings, the office offers cake or ice cream during the all-hands meetings and afterwork events once a month. Admittedly, if all would show up in the office to work on a regular day, everybody would no longer have a desk with a large screen, but there is still room for everybody due to a variety of seating solutions.

### Company-Y

Company-Y is a multinational conglomerate corporation that employs over 100'000 employees worldwide. We studied one Swedish site of Company-Y with around 500 employees.

The post-pandemic hybrid work policy in the company site in Sweden permits WFH two-three days per week. Office presence is seen by the management as a catalyst for collaborative, problem-solving and creative work, as well as efficient decision making.

The initial return to the office at Company-Y was slow, which motivated mandatory office presence. Yet, average weekly office presence does not exceed 50%. One potential factor that perhaps hindered commute was the cancellation of the free parking and free charging stations for electrical vehicles. However, generally, remote work is associated more with the focus on individual preferences.

Three major actions were taken to change this attitude. First, to increase the density of employees in the office and make the environment seem more social, Company-Y downsized the office space. This action was also followed by renovations, making the offices cozier and improving the local canteen. Finally, Company-Y cultivates "We over I" culture and coach teams to have team days with co-presence.

### SpareBank 1 Utvikling

SB1U is a Norwegian software company owned by an alliance of banks that employs 25 software teams developing and maintaining the top-rated mobile banking application in Norway.

Office presence at SB1U is seen as the key to innovation, as well as the prerequisite for maintaining efficiency, cooperation, interaction in the teams and a social environment in general. Starting from July 2022, a new national regulation mandates that employees and employers must establish a written agreement outlining the extent to which remote work is allowed. Following extensive internal deliberation, SBU1 decided on the 50% minimum office presence. Due to extra capacity in the office as people worked from home, SB1U decided not to expand the office space when hiring 100 more people. This meant that most could not have their own desk anymore and on days with high office presence, some could end up without a desk.

To balance individual, team, and organizational needs in the new flexible work life, SB1U allocated fixed zones for teams. This allowed in the absence of personal desks and fewer seats still manage to facilitate everyone. After all, someone is always away, and seats can be borrowed nearby. To increase office attractiveness, SBU1 also improved bicycle parking and organizes onsite-only meetings, afterwork events, and invites pop-up barista.

## Spotify

Spotify is a major music streaming service provider employing almost 10'000 employees, of which roughly 1/5 are engineers who develop and maintain digital music, podcast, and video services. Spotify has five main R&D sites, of which our study included two – in Stockholm and London.

After the pandemic, Spotify has announced the new "Work from Anywhere" policy, permitting employees to choose where to work from upon manager's approval, including the possibility for international relocation. The company intention is to become remote-first. This means that the main purpose of the office is to support those who find it problematic to work remotely and provide socialization opportunities and space for occasional onsite activities.

The main challenges at Spotify, in both locations, include low office occupancy. The pre-pandemic office capacity coupled with very low office presence resulted in offices feeling empty, which decreases the motivation to commute to the office for those interested in socialization and collaboration. Further, London-based employees often belong to team based in Sweden, and thus do not have teammates locally.

To address these challenges, Spotify sites organize onsite events for employees across different departments, and offer a variety of office perks, starting with very central office locations, and followed by numerous socialization spaces (whole floors), free snacks and beverages, fulltime barista, free lunches in the UK.

## Storebrand

Storebrand is a Norwegian financial services company with over 2'000 employees. The company operates in the Nordic markets and delivers pension, savings, insurance and banking products to individuals, businesses, and public enterprises. Our study focused on one Norwegian site of Storebrand with 1500 employees.

To satisfy individual needs and maintain low retention after the pandemic, Storebrand introduced a very flexible work policy with only one exception – fully remote is not an option. Storebrand laid its strategy already before the end of the pandemic. The company intentionally fosters discussions over rules to build habits around sound reasoning and co-operation rather than strict policies.

Hybrid strategy at Storebrand was based on a belief that a one-size-fits-all solution does not exist and even one, lasting solution for the same team can be utopic. Instead, management focuses on team-based flexibility, and promotes regular discussions of the need for and importance of co-presence. Besides, Storebrand introduced several measures to revamp the office. At the end of the pandemic, the office was renovated to reflect the changes that happened during the two years of forced WFH. Several pilots were initiated, including dedicated focus area called "the library" and bookable team areas to ensure close seating on the co-presence days. "Company Tuesdays" was launched as a co-presence day featuring work-related events, knowledge sharing forums and office perks (mostly food and drinks). This was later pivoted to "Happy Fridays", not to attract more people on Fridays, but to make Fridays a good day for those who chose to come. The company plans to pivot the idea again, focusing on Mondays. All these efforts, notably without mandatory office days, made Storebrand succeed with achieving the highest office occupancy among the companies in our study.

## Telenor

Telenor is one of the leading telecommunications companies headquartered in Norway with around 16'000 employees. The company owns and operates mobile networks in Scandinavian and several Asian countries. In our study, we included a smaller Telenor site in Karlskrona, Sweden, and a large site in Oslo, Norway.

Remote work at Telenor is not centrally regulated and each country has their own local regulations. In Norway, Telenor employees are 3'000 In Sweden, employees are expected to work at least 2 days onsite. At the same time, office is promoted as the place to cultivate company culture, loyalty and team building.

Despite the rules for mandatory presence, in many departments, office presence is not strictly enforced, leading to having employees who work entirely remotely. Due to relatively low office presence, the role of the office is also a reoccurring topic of discussion.

Telenor puts emphasis on the collaborative culture and teamwork, supported by the slogan "We go before Ego". Another ongoing initiative to make Telenor offices economical and attractive, is office renovation, which has already started. The renewed offices will have focus areas, phone booth, meeting rooms, recreation and socialization areas, and be based on a free sitting.

| | | Ericsson | Knowit | Company X | Company Y | SBU1 | Spotify | Spotify | Storebrand | Telenor | Telenor |
|---|---|---|---|---|---|---|---|---|---|---|---|
| | | 🇸🇪 | 🇳🇴 | 🇸🇪 | 🇸🇪 | 🇳🇴 | 🇸🇪 | 🇬🇧 | 🇳🇴 | 🇸🇪 | 🇳🇴 |
| | | Karlskrona | Oslo | Malmö | Lund | Oslo | Stockholm | London | Oslo | Karlskrona | Oslo |
| **Policy** | Attitude | Remote-friendly | Flexible | Remote-friendly | Remote-friendly | Remote-friendly | Flexible | Flexible | Flexible | Remote-friendly | Flexible |
| | Permitted amount of remote work | 50:50% in a year | Flexible, negotiated | | 2-3 days/week | 50:50% | Up to 100% | Up to 100% | Fully remote is not an option | Max 3 days/week | |
| **Occupancy** | Weekly average | 22 | Around 70% | 30% | 44% | 47% | N/A | N/A | 55% | N/A | N/A |
| | Min (weekly) | 25 | Below 70% | 24 | 39% | 46% | N/A | N/A | 54% | N/A | N/A |
| | Max (weekly) | 19 | Around 80% | 37% | 47% | 48% | N/A | N/A | 55% | N/A | N/A |
| | Weekday average | 18 26 24 15 | | 34 36 30 32 18 | 36 55 55 45 29 | 42 59 60 50 25 | N/A | N/A | 52 65 64 62 35 | 58 58 63 46 23 | N/A |
| **Perks** | Food and drinks | Free coffee, tea | Free coffee, tea | Free coffee, tea Free fruit, Occasion. treats (ice cream, cake) | Free coffee, tea Free fruit | Free coffee, tea | Barista coffee Free snacks and beverages | Barista coffee Free snacks, beverages, lunch | Free coffee, tea Free fruit | Free coffee, tea Occasion. treats (buns, candy) | Free coffee, tea |
| | Office activities | | | | Gym | Gym, table tennis, shuffleboards | Various sport & video games | | Gym and sport arena | | Table tennis |
| | View | | | | Spectacular view | | | | | Spectacular view | |
| | Location | Central, harbour | Central | Suburbs, nice area | Suburbs | Central | Central | Central | Suburbs, nice area | Nice area, seafront | Suburbs, seafront |
| **Changes** | Renovations | Hybrid-friendly meeting rooms, canteen made open to public | New, better office, Many social areas | New, better office | Renovated office, WFH zones, new socialization zones, better canteen | Fixed team areas; Open office | New socialization areas with sofas, Quite zones | New socialization areas with sofas, Quite zones | More space, new colors, plants, new desk setup, newly equipped meeting rooms, future office pilots | (Major renovation plans) | |
| | Size | | Fewer desks | | Downsized | | Fewer desks | Fewer desks | | (Downsizing plan) | |
| | Parking | Limited places | Just bicycle parking | | No free parking | Better bicycle parking | | | | | |
| | Tool support | | Booking system | | Booking system | Booking system | | | | | |
| **Activity spaces** | Formal events | Group "return-to-office" events, seminars | Subject group meetings, Monthly unit meetings | All-hands meetings | All-hands meetings, breakfast seminars | Some onsite-only meetings, tech conference | | | | Top managers in the office; Seminars | Lunch seminars |
| | Informal events | Afterwork events, team breakfasts | Waffles and beer on Fridays | Monthly afterwork | Regular afterwork | Afterwork events, pop-up barista | Musicians as guests | Musicians as guests | Company Tuesday, Happy Friday, afterwork | Afterwork events | |

Office occupancy heatmap scale: 1-10  11-20  21-30  31-40  41-50  51-60  61-70  71-80  81-90  99-100

**Table 2**: Overview of the companies and their offices, attractions, perks and activities that impact office presence

## What brings employees to the office?

Our analysis of self-reported reasons for office presence indirectly validates which office attractions matter, grouped into four major categories (see Figure 1): Connection and community, Material offerings, Preference (for working in the office), and Duty. Interestingly, we found that motivation can be positive – reasons to come to the office (for example, a good workplace in the office), and negative – reasons not to work from home (for example, a poor workplace at home).

The most common reason for onsite work is related to Connection and community (101 votes), i.e., to socialize and collaborate. Respondents in this category report practical reasons to be onsite (meeting people, teamwork) along with emotional reasons (energy, spontaneity, informality, and peer support).

The second important category of reasons to visit the office comprises material offerings related to the workspace, location as well as office perks. Respondents in this category, report being drawn to the office by good coffee, better lunch options, especially free food, better physical workspace, and office location. Some respondents come in when having errands around the office. Interestingly, while material reasons may seem entirely practical, we find emotional reasons here too (waffles and wonderful view).

In the third group, we find general preferences for office work, such as increased productivity, separation of work and personal life and avoidance of distractions at home. For some, in this category, productivity is associated with physical office conditions, while for others it is more emotional – the stimulating atmosphere and social pressure.

Finally, the fourth group comprises duties that draw respondents to work onsite (appointments), the general sense of duty, mandatory presence and access to physical infrastructure.

The mentioned reasons surely are company dependent. For example, not every company offers barista coffee or a spectacular sea view. Similarly, not all companies promote teamwork and onsite activities. However, we believe that the aggregated list of reasons provides a good indication of what matters in the eyes of employees.

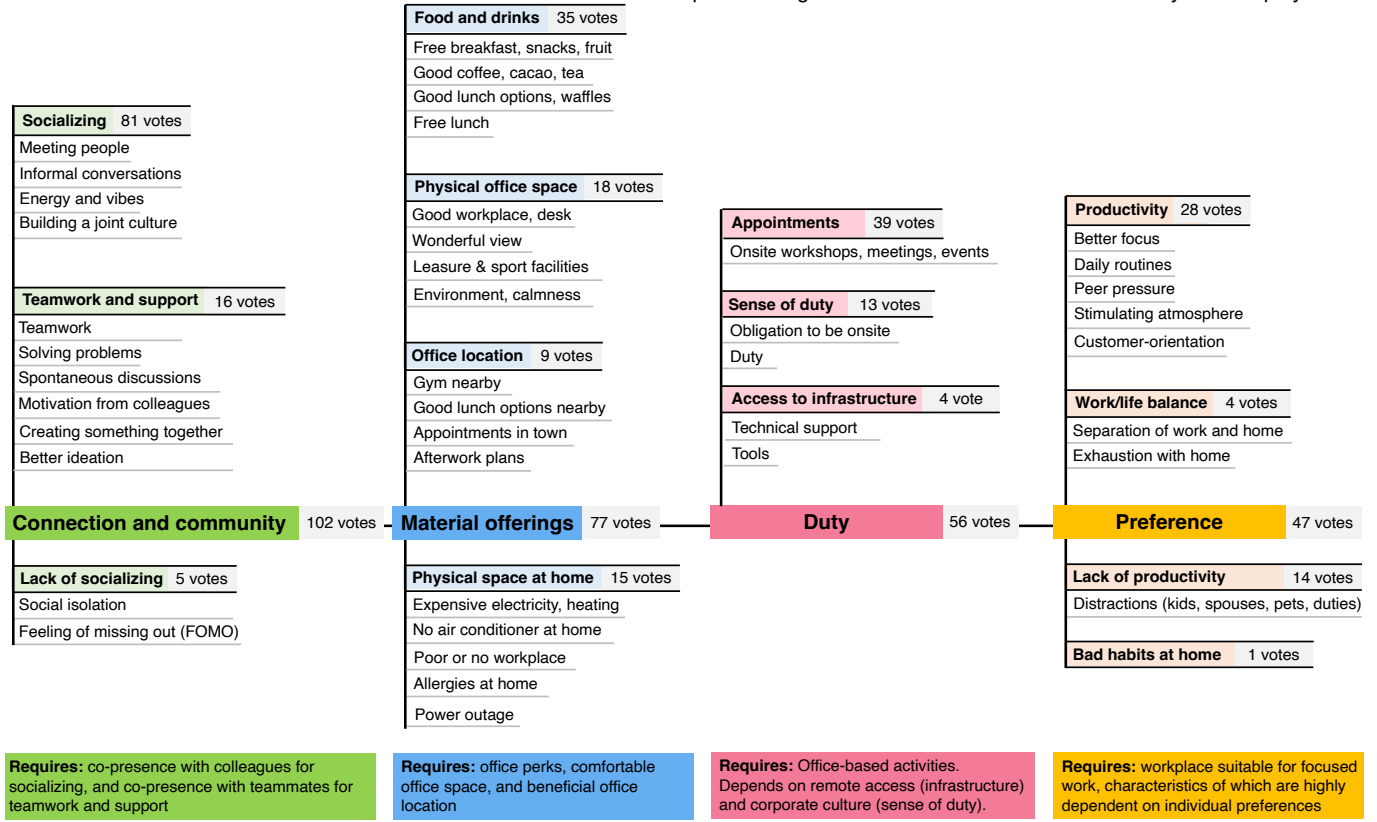

**Figure 1:** The main motivations for office presence ordered by importance (based on the number of supporting votes). The upper part of the graph depicts positive motivation (reasons to come to the office) while the bottom part depicts negative motivation (reasons not to WFH).

## Practical implications

Readers may reasonably ask whether office presence matters. Is there anything that we today cannot achieve through digital interaction? Our results show that office still plays a role as the source of energy, strong peer support, and the place for building joint culture. We also found that some employees simply prefer onsite work over WFH. The other significant motivations for office presence include Connection and community and Material offerings, which are known corporate value propositions stimulating positive experiences at work.

**Connection and community** is the most prominent motivation for office presence. Research shows that these contribute to the sense of belonging, which not only makes employees to want to work together, but also helps to retain employees in long term [9]. To foster connection and community, companies shall cultivate good relationships and offer an energizing company culture. Obviously, these are hard to achieve when everybody or too many work remotely. For policy makers and unit managers, we suggest:

> Stimulate co-presence through corporate onsites or mutual agreements in the teams and work groups.

> Support formation of interest groups by providing onsite facilities (gym, video-gaming), equipment (table tennis) or paying for corporate offsite classes and memberships

> Organize onsite activities, formal (seminars, regular meetings) and informal (afterwork events, special guests)

And for employees working in teams, we remind that one's absence ultimately influences the socialization and collaboration experiences of others. Efficient teamwork requires team-orientation [14]. We thus suggest:

> Consider team needs over individual needs

Although **material offerings** might not provide lasting experiences or have a prominent impact on retention [9], we found that office perks were not unimportant. Many reasons in this category served as hygiene factors [10] such as dissatisfaction with home environment (distractions, poor or no workplace, expensive electricity, no air conditioner) or factors that were superior at work (good coffee / lunch options, wonderful view). Fairly, those who upgrade home office or buy a better coffee machine might no longer be motivated to work onsite. A more permanent factor is the office location. We learned that central locations close to many important municipal and entertainment services helps to increase office presence. For facility managers, we suggest:

> Ensure an attractive and comfortable workplace for employees who prefer onsite work

> Optimize office place for teams to facilitate collaboration and socialization

> Ensure good food/drink options at the workplace, and lunch options; free perks are of extra benefit

> Locate the office strategically, close to municipal and entertainment services and within ease of reach

## About the Authors

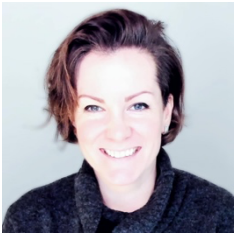

**Darja Šmite** is a professor of software engineering at the Blekinge Institute of Technology in Sweden, where she leads research efforts on remote work, distributed software development, and outsourcing. Her research interests include WFH, virtual and hybrid teams, large-scale agile, team performance, autonomy, and organizational decentralisation. Šmite received a PhD in Computer Science from University of Latvia. She is a part time research scientist at SINTEF in Norway, prior to her academic career, Šmite has worked as a software engineer, systems analyst and IT consultant in Latvia. Contact her at darja.smite@bth.se.

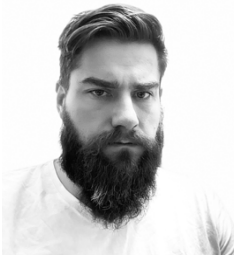

**Eriks Klotins** is a senior researcher at the Blekinge Institute of Technology in Sweden. He earned his PhD from the same institute in 2019 on Software Engineering practices in start-up companies. Before joining academia, Eriks had extensive experience in various engineering roles. Since 2020, his primary focus has been on continuous software engineering. Specifically, how large organizations can streamline their software delivery processes and practices to maximize internal efficiency, customer value delivery, and sustainability. Contact him at eriks.klotins@bth.se

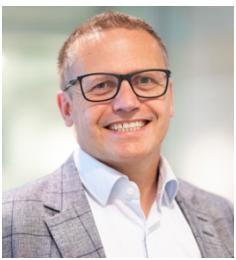

**Nils Brede Moe** is a chief scientist at SINTEF in Norway. He works with software process improvement, intellectual capita, innovation, autonomous teams, agile and global software development, and digital transformation. He has led several nationally funded software engineering research projects covering organizational, sociotechnical, and global/distributed aspects. Moe received a Dr. Philos. in Computer Science from the Norwegian University of Science and Technology and holds an adjunct position at the Blekinge Institute of Technology in Sweden. Contact him at nils.b.moe@sintef.no